\documentclass[conference,10pt]{IEEEtran}
\IEEEoverridecommandlockouts
\usepackage{amsmath,amssymb,amsfonts}
\usepackage[shortlabels]{enumitem}
\usepackage[lined,ruled,linesnumbered]{algorithm2e}
\SetKwComment{Comment}{/* }{ */}
\usepackage{algorithmic}
\usepackage{graphicx}
\usepackage{textcomp}
\usepackage{xcolor}
\usepackage{bm}
\usepackage{epsfig}
\usepackage{subfigure}
\usepackage{psfrag}
\usepackage{cite}

\def\BibTeX{{\rm B\kern-.05em{\sc i\kern-.025em b}\kern-.08em
		T\kern-.1667em\lower.7ex\hbox{E}\kern-.125emX}}

\begin{document}
\title{{\color{black} AirFL-Mem: Improving Communication-Learning Trade-Off by Long-Term Memory}}
\author{\IEEEauthorblockN{Haifeng Wen${}^{\ast}$, Hong Xing${}^{\ast\S}$, and Osvaldo Simeone${}^\dagger$}\\
	\IEEEauthorblockA{${}^\ast$ The Hong Kong University of Science and Technology (Guangzhou), Guangzhou, China \\
		${}^\S$ The Hong Kong University of Science and Technology, HK SAR, China \\
        ${}^\dagger$ KCLIP Lab, Centre for Intelligent Information Processing Systems (CIIPS), Department of Engineering,\\ King's College London, London, U.K.\\
		E-mails:~hwen904@connect.hkust-gz.edu.cn,~hongxing@ust.hk.,~osvaldo.simeone@kcl.ac.uk}
 \thanks{ The work of  O. Simeone was supported by the European Union’s Horizon
Europe project CENTRIC (101096379), by an Open Fellowships of the EPSRC
(EP/W024101/1), by the EPSRC project (EP/X011852/1), and by Project REASON, a UK Government funded project.}
}

\maketitle

\begin{abstract}
Addressing the communication bottleneck inherent in federated learning (FL), over-the-air FL (AirFL) has emerged as a promising solution, which is, however, hampered by  deep fading conditions. In this paper, we propose AirFL-Mem, a novel scheme designed to mitigate the impact of deep fading by implementing a \emph{long-term} memory mechanism. Convergence bounds are provided that account for long-term memory, as well as for existing AirFL variants with short-term memory, for general non-convex objectives. The theory demonstrates that AirFL-Mem exhibits the same convergence rate of federated averaging (FedAvg) with ideal communication, while the performance of existing schemes is generally limited by error floors. The theoretical results are also leveraged to propose a novel convex optimization strategy for the truncation threshold used for power control in the presence of Rayleigh fading channels. Experimental results validate the analysis, confirming the advantages of a long-term memory mechanism for the mitigation of deep fading.
\end{abstract}


\IEEEpeerreviewmaketitle

\newtheorem{definition}{\underline{Definition}}[section]
\newtheorem{fact}{Fact}
\newtheorem{assumption}{Assumption}
\newtheorem{theorem}{\underline{Theorem}\underline{\hspace{1.5em}}\hspace{-1.5em}}[section]
\newtheorem{lemma}{\underline{Lemma}}[section]
\newtheorem{proposition}{\underline{Proposition}}[section]
\newtheorem{corollary}[proposition]{\underline{Corollary}}
\newtheorem{example}{\underline{Example}}[section]
\newtheorem{remark}{\underline{Remark}}[section]
\newcommand{\mv}[1]{\mbox{\boldmath{$ #1 $}}}
\newcommand{\mb}[1]{\mathbb{#1}}
\newcommand{\Myfrac}[2]{\ensuremath{#1\mathord{\left/\right.\kern-\nulldelimiterspace}#2}}
\newcommand\Perms[2]{\tensor[^{#2}]P{_{#1}}}

\section{Introduction} \label{sec:Introduction} 


Over-the-air FL (AirFL) has emerged from information-theoretic studies \cite{nazer2007computation} as a promising approach to enable model aggregation in wireless implementations of federated learning (FL) \cite{zhu2019broadband}. A well-known problem with AirFL is that devices experience different  fading conditions, causing the aggregated model estimated by the central server to deviate from the desired model average, unless strict power constraints mechanisms are applied. 

In the presence of channel state information at the transmitter (CSIT), the typical solution to this problem   is to ensure signal alignment through truncated channel inversion \cite{zhu2019broadband,sery2021over,cao2021optimized,shah2022robust, yao2023imperfect,amiri2020federated}. Truncation entails that only a subset of model parameters reach the server, causing the erasure of potentially important information. Reference \cite{amiri2020federated} proposed to mitigate the resulting channel-driven sparsification of model information  via error feedback. Specifically, the approach therein applies a short-term memory mechanism that operates across two successive iterations. However, no theoretical guarantees are currently available for this mitigation strategy. This paper aims at addressing this knowledge gap, revealing through theoretical bounds that error feedback based on longer-term memory mechanisms is generally necessary to combat the effect of deep fading on truncated channel inversion.
 
To provide additional context for this work, other solutions to the problem of deep fading in AirFL in the presence of CSIT include phase-only compensation \cite{sery2020analog,paul2021accelerated}, transmission weight optimization \cite{yang2020federated}, and device selection \cite{xia2021fast}. 
Without CSIT, references \cite{amiri2021blind, yang2021revisiting, wei2023random, tegin2023federated} studied the convergence of AirFL over broadband channels, and improved the performance by exploiting the channel hardening property of massive multiple-input and multiple-output (MIMO) channels. Error-feedback-based transmission has been widely adopted for communication-efficient FL with the aim of compensating losses for model-update information. Such methods compensate for accumulated errors due to artificially induced sparsity or quantization in the next-round transmission \cite{amiri2020federated, stich2018sparsified, basu19qsparse, karimireddy2019error, yun2023communication}. For digital transmission-based FL, the authors of \cite{yun2023communication} applied a memory vector in digital FL schemes to compensate for both compression and reconstruction errors.

Overall, in this paper, inspired by sparsified stochastic gradient descent (SGD) with memory \cite{stich2018sparsified}, we introduce AirFL-Mem, an AirFL protocol that  implements a long-term memory mechanism for error feedback in truncated channel inversion. Furthermore, we analyze the role of memory in error feedback via convergence bounds. 
Our contributions are as follows:
\begin{itemize}
    \item We provide convergence bounds for AirFL-Mem, as well as for the existing variant with short-term memory \cite{amiri2020federated}, demonstrating that AirFL-Mem achieves the same convergence rate as FedAvg in perfect communication conditions \cite{mcmahan2017communication}, while a shorter memory may cause an error floor.
    \item Based on the derived bounds, we introduce a novel convex optimization-based truncation-threshold selection scheme for the implementation of AirFL-Mem.
\end{itemize}

The rest of the paper is organized as follows. Sec. II introduces system level and preliminaries. Sec. III describes AirFL-Mem, while Sec. IV describes the derived theoretical bounds. Sec. V presents the proposed optimization scheme for the power control thresholds, and Sec. VI covers numerical results, with Sec. VII completing the paper.

\section{System Model and Preliminaries} \label{sec:System Model} 
In this section, we consider an AirFL system in which a set  $[K] \triangleq \{1, \ldots, K\}$ of devices transmit their machine learning models to an edge server over a Gaussian multiple access channel (MAC) fading channel. In this section, we first introduce the vanilla FL protocol premised on ideal and noiseless communications, and then describe the considered communication model accounting for fading and noise.
\subsection{Learning Protocol (Vanilla FL)} \label{subsec:Learning Model}
In the FL setup, each device $k \in [K]$, possesses a distinct local dataset denoted as $\mathcal{D}_{k}$. All devices share a machine learning model, e.g., a neural network or a transformer parameterized by a vector $\mv{\theta} \in \mathbb{R}^{d \times 1}$. The objective of the FL system is to collaboratively solve the empirical loss minimization problem
\begin{align*}
\mathrm{(P0)}:&~\mathop{\mathtt{Minimize}}_{\mv \theta}~~~f(\mv \theta)\triangleq\frac{1}{K}\sum_{k=1}^{K}f_{k}(\mv \theta),
\end{align*} 
where $f(\mv \theta)$ represents the global empirical loss function; and $f_{k}(\mv \theta) = 1/\vert\mathcal{D}_{k}\vert\sum_{\bm\xi\in\mathcal{D}_{k}}\mathcal{L}(\mv \theta;\bm\xi)$ represents the local empirical loss function for device $k \in [K]$. The cardinality operation $|\cdot|$ denotes the size of a given set, and the notation $\mathcal{L}(\mv{\theta};\bm{\xi})$ indicates the loss function evaluated at parameter $\mv{\theta}$ with respect to (w.r.t) the data sample $\bm{\xi}$.

In the following, we briefly review the standard \emph{FedAvg} protocol \cite{mcmahan2017communication}. Let \(t\in\{0,\ldots, T-1\}\) denote the index of global communication rounds or, equivalently, of global iterations. At the $t$-th global iteration, individual devices obtain a localized parameter $\mv \theta_{k}^{(t)}$ that approximates the solution \(\mv \theta^{(t)}\) to problem ($\mathtt{P0}$) by minimizing the local loss $f_{k}(\mv \theta)$. Specifically, during each global iteration $t$, each device executes $Q$ local SGD steps over its private dataset $\mathcal{D}_{k}$, leading to the iterative update of the local model parameter $\mv \theta_{k}^{(t)}$ given by
\begin{align}
\mv \theta_{k}^{(t, q+1)} \leftarrow \mv \theta_{k}^{(t, q)}-\eta^{(t)}\hat\nabla f_{k}(\mv \theta_{k}^{(t, q)}), \label{eq:local updates}
\end{align} 
where $q\in\{0,\ldots, Q-1\}$ is the \emph{local iteration} index; $\eta^{(t)}$ denotes the learning rate; and $\hat\nabla f_{k}(\mv \theta_{k}^{(t,q)})$ is the estimate of the true gradient $\nabla f_{k}(\mv \theta_{k}^{(t,q)})$ computed from a mini-batch $\mathcal{D}_{k}^{(t)}\subseteq\mathcal{D}_{k}$ of data samples, i.e.,
\begin{align}
\hat\nabla f_{k}(\mv \theta_{k}^{(t, q)})=\frac{1}{\vert\mathcal{D}_{k}^{(t)}\vert}\sum\limits_{\bm\xi\in\mathcal{D}_{k}^{(t)}}\nabla \mathcal{L}(\mv \theta_{k}^{(t, q)};\bm\xi). \label{eq:SGD}
\end{align} 
The initialization of local model parameters commences with the shared model parameter held by the edge server as \(\mv \theta_{k}^{(t,0)}=\mv \theta_{k}^{(t)}\leftarrow \mv \theta^{(t)}\).

At each global iteration $t$, the edge server receives the \emph{model differences} \begin{equation}\label{eq:modeldiff}\mv{\Delta}_{k}^{(t)} = \mv \theta_{k}^{(t,0)}-\mv \theta_{k}^{(t, Q)}, \end{equation} $k \in [K]$. Subsequently, it aggregates these differences to update the global model parameters as
\begin{equation}
    \mv \theta^{(t+1)} \leftarrow \mv \theta^{(t)} - \frac{1}{K}\sum_{k=1}^{K}\mv{\Delta}_{k}^{(t)}. \label{eq:global updates}
\end{equation} 
This updated model parameter is then broadcast to all $K$ devices, serving as the initialization of their local iterates (c.f. \eqref{eq:local updates}). The above steps are iterated until some convergence criterion is met.

\subsection{Communication Model} \label{subsec:Communication Model}

At the $t$-th global communication round, each device $k \in [K]$ uses orthogonal frequency division multiplexing (OFDM) to transmit each entry of the scaled model difference $\mv \Delta_{k}^{(t)}$ in \eqref{eq:modeldiff} over one of a total of $s$ subcarriers for $\lceil{\frac{d}{s} \rceil}$ OFDM symbols. Specifically, the $d\times 1$ vector of transmitted signal is given by  $\mv x_{k}^{(t)} = \mv \Delta_{k}^{(t)}/\eta^{(t)}$ \cite{zhu2019broadband, amiri2020federated}. 
With $x_{k,j}^{(t)}$ denoting the $j$-th entry of vector $\mv x_{k}^{(t)}\in \mathbb{R}^{d\times 1}$, the edge server receives the signal simultaneously transmitted by all $K$ devices as
\begin{equation}
    y_{j}^{(t)}=\sum_{k=1}^{K}\sqrt{\kappa_{k}}h_{k,j}^{(t)}p_{k,j}^{(t)}q_{k,j}^{(t)}x_{k,j}^{(t)}+n_{j}^{(t)},
    \label{eq:rxsignal}
\end{equation}
where $\kappa_{k}$ denotes the large-scale fading induced channel gain between device $k$ and the edge server; $h_{k,j}^{(t)}$ is the small-scale fading coefficient affecting the $j$-th entry at the $t$-th communication round; $p_{k,j}^{(t)}$ represents a power control factor and $q_{k,j}^{(t)}\in \{0,1\}$ binary masking variable, both of which with be explained next; $n_{j}^{(t)}$ is the independent and identically distributed (i.i.d.) additive Gaussian noise with zero mean and variance $\sigma^2$. Each device $k$ has per-channel use power constraint $P_{k}$.

Under the assumption of perfect CSIT, truncated channel inversion transmission ensures that model parameters transmitted by all active devices are effectively aligned at the receiver \cite{zhu2019broadband}.
Accordingly, each device $k$ transmits entry $x_{k,j}^{(t)}$ only if the corresponding channel gain $|h^{(t)}_{k,j} |^2$ is larger than a threshold $\epsilon_{k}>0$. This is ensured by choosing the masking variable as
\begin{equation}
    q^{(t)}_{k,j}=
    \begin{cases} 1, & |h^{(t)}_{k,j} |^2 \ge \epsilon_{k} \\ 0 & \text{otherwise} \end{cases}
    \label{eq:indicator variable}
\end{equation}
for some device-specific thresholds $\epsilon_{k}$.
Furthermore, the power scaling factor $p_{k,j}^{(t)}$, $k \in [K]$, is set as
\begin{equation}
    p^{(t)}_{k,j}=\frac{\sqrt{\rho^{(t)}}}{\sqrt{\kappa_{k}} h_{k,j}^{(t)}}, 
    \label{eq:power scaling factor}
\end{equation}
where the common scaling factor $\rho^{(t)}$ is selected to guarantee a per-block power constraint at communication round $t$. Denoting $\mv p_{k}^{(t)}=[p^{(t)}_{k,1},p^{(t)}_{k,2},\cdots,p^{(t)}_{k,d}]^T$ and $\mv q_{k}^{(t)}=[q^{(t)}_{k,1},q^{(t)}_{k,2},\cdots,q^{(t)}_{k,d}]^T$ as the power scaling and the masking vectors, respectively, the power constraint can be expressed as 
{\color{black}
\begin{equation} \label{eq:power constraint}
    \frac{1}{d} \mathbb{E}\| \mv p_{k}^{(t)} \odot \mv q_{k}^{(t)} \odot \boldsymbol x_{k}^{(t)} \|^2 \le P_{k}
\end{equation}
}
for all $k \in [K]$, where $\odot$ is the element-wise product.

Finally, upon receiving the vector $\mv y_{k}^{(t)}=[y_{k,1}^{(t)},y_{k,2}^{(t)},\cdots,y_{k,d}^{(t)}]^T$, the edge server estimates the aggregated model differences and performs global update as
\begin{equation}
    \mv \theta^{(t+1)}=\mv \theta^{(t)}-\frac{\eta^{(t)}}{\sqrt{\rho^{(t)}}K} \mv y^{(t)}.
    \label{eq:airfl global update}
\end{equation}


\section{AirFL with Long-Term Memory (AirFL-Mem)} \label{sec:AirFLMem}

\begin{algorithm}[t] \label{alg:Algorithm 1}
\SetKwInOut{Input}{Input}
\SetKwInOut{Output}{Output}
\SetKwBlock{DeviceParallel}{On devices $k \in [K]$:}{end}
\SetKwBlock{localSGD}{for $q=0$ to $Q-1$ do}{end}
\SetKwBlock{OnServer}{On Server:}{end}
\caption{AirFL-Mem}\label{alg:Algorhtm}
\textbf{Input:} learning rate $\eta^{(t)}$, power constraints $\{P_{k}\}_{k \in [K]}$, number of global rounds $T$, number of local rounds $Q$ \\
Initialize $\mv \theta_{k}^{(0)}=\mv \theta^{(0)}$ and $~\mv m_{k}^{(0)} = \mv 0$ for all $k \in [K]$ and $t=0$ \\
\While{$t < T$}{
\DeviceParallel{
$\mv{\theta}^{(t,0)}_{k}\leftarrow \mv{\theta}^{(t)}$\;
\localSGD{
$\mv{\theta}^{(t,q+1)}_{k}\leftarrow \mv{\theta}^{(t,q)}_{k} - \eta^{(t)} \hat\nabla f(\mv \theta_{k}^{(t,q)})$
}
Calculate model difference $\mv{\Delta}_{k}^{(t)} = \mv \theta_{k}^{(t,0)}-\mv \theta_{k}^{(t, Q)}$ \\
Update long-term memory variable $\mv{m}_{k}^{(t+1)}=\boldsymbol m^{(t)}_{k}+\boldsymbol \Delta^{(t)}_{k} - \mv q^{(t)}_{k}\odot(\boldsymbol m^{(t)}_{k}+\boldsymbol \Delta^{(t)}_{k})$ \\
Transmit $\mv x_{k}^{(t)} = (\boldsymbol m^{(t)}_{k}+\boldsymbol \Delta^{(t)}_{k})/\eta^{(t)}$ \\
}
\OnServer{
Receive $ y_{j}^{(t)}=\sum_{k=1}^{K}\sqrt{\kappa_{k}}h_{k,j}^{(t)}p_{k,j}^{(t)}q_{k,j}^{(t)}x_{k,j}^{(t)}+n_{j}^{(t)}$, for $m\in[d]$\\
Global Update: $\mv \theta^{(t+1)}=\mv \theta^{(t)}-\frac{\eta^{(t)}}{\sqrt{\rho^{(t)}}K} \mv y^{(t)} $\\
Broadcast $\mv \theta^{(t+1)}$ to all $K$ devices\\
}
$t \leftarrow t + 1$\\
}
\end{algorithm}

In this section, we propose a new AirFL scheme, referred to as \emph{AirFL-Mem}, that utilizes long-term memory to mitigate the learning performance loss due to masking caused by deep fading.

\subsection{Motivation} \label{subsec:Motivation}
When channel conditions are poor for device $k$, the channel gain may satisfy the inequality $|h^{(t)}_{k,j} |^2 < \epsilon_{k}$. By \eqref{eq:rxsignal} and \eqref{eq:indicator variable}, this causes the masking of the corresponding model difference $\Delta_{k,j}^{(t)}$ encoded by signal $x_{k,j}^{(t)}$. 
The discrepancy between the actual model difference vector $\mv \Delta_{k}^{(t)}$ and its truncated version $\mv q_{k}^{(t)} \odot \mv \Delta_{k}^{(t)}$ is given by $(\mv 1-\mv q_{k}^{(t)})\odot \mv \Delta_{k}^{(t)}$, where $\mv 1$ is the $d \times 1$ all-one vector.
In prior art, the authors in \cite{zhu2019broadband} treated this error as a form of ``dropout'', and they did not perform any compensation. In contrast, a \emph{short-term} memory mechanism was introduced in \cite{amiri2020federated}. In it, at the current round $t$, device $k$ transmits the compensated signal 
\begin{equation}
    \mv x_{k}^{(t)} = ( \mv \Delta_{k}^{(t)}+\tilde{\mv m}_{k}^{(t)} )/\eta^{(t)},
    \label{eq:transmitted signal of ECESA-DSGD}
\end{equation}
where $\tilde{\mv m}_{k}^{(t)}=(\mv 1 - \mv q_{k}^{(t-1)}) \odot \mv \Delta_{k}^{(t-1)}$, with $\tilde{\mv m}_{k}^{(0)}=\mv 0$, accounts for the masking-induced discrepancy at the previous round.
We will show in Sec. \ref{sec:main results} that these two schemes lead to an error floor in terms of convergence to stationary point, and that a long-term memory mechanism can mitigate this problem.

\subsection{AirFL-Mem}
Inspired by error feedback schemes that are widely adopted to mitigate sparsity-induced errors \cite{stich2018sparsified, basu19qsparse, karimireddy2019error}, we propose a \emph{long-term} memory mechanism to partially compensate for the distortion caused by masking due to the power control policy \eqref{eq:indicator variable}. To this end, each device $k \in [K]$ accumulates a long-term error $\mv m_{k}^{(t)}$ variable
\begin{equation}
    \mv{m}_{k}^{(t+1)}=\mv{m}_{k}^{(t)}+\mv e_{k}^{(t)},
    \label{eq:memory update rule}
\end{equation}
where $\mv{m}_{k}^{(0)}=\mv 0$ and the discrepancy $\mv e_{k}^{(t)} \in \mathbb{R}^{d\times 1}$ at round $t$ is given by 
\begin{equation}
 \mv e_{k}^{(t)}=\boldsymbol \Delta^{(t)}_{k}-\boldsymbol q^{(t)}_{k}\odot (\boldsymbol \Delta^{(t)}_{k}+\boldsymbol m^{(t)}_{k}).
 \label{eq:discrepancy with memory}
\end{equation}
In \eqref{eq:discrepancy with memory}, the current model difference $\mv \Delta_{k}^{(t)}$ is corrected by the accumulated error $\mv m_{k}^{(t)}$ prior to the application of masking.
Accordingly, each device $k \in [K]$ transmits the compensated model difference
 \begin{equation}
     \mv x_{k}^{(t)} = (\boldsymbol \Delta^{(t)}_{k} + \boldsymbol m^{(t)}_{k})/\eta^{(t)}.
     \label{eq:transmit signal with memory}
 \end{equation}
As compared to \cite{amiri2020federated}, which compensated only the error due to the previous round of communication, the long-term error variable $\mv{m}_{k}^{(t)}$ in \eqref{eq:memory update rule} accounts for the error accumulated from the beginning up to the current round of training.
The proposed AirFL-Mem is summarized in Algorithm \ref{alg:Algorithm 1}.


\section{Convergence analysis of AirFL-Mem} \label{sec:main results}
In this section, we study the convergence performance of the proposed AirFL-Mem in general non-convex settings.
The main goal is to obtain insights into the role of the long-term memory compensation mechanism \eqref{eq:memory update rule}-\eqref{eq:discrepancy with memory} in the convergence of AirFL protocols.
\subsection{Assumptions} \label{subsec:assumptions}
We make the following assumptions.
\begin{assumption}[$L$-smoothness] \label{assumption: L-smoothness}
    For all $k \in [K]$, the local empirical loss function $f_{k}$ is differentiable and $L_{k}$-smooth, i.e., $ \text{for all } \mv w, \mv u \in \mathbb{R}^d$, we have the inequality
    \begin{equation}
        \left\|\nabla f_{k}(\mv w)-\nabla f_{k}(\mv u)\right\| \leq L_{k}\|\mv w-\mv u\|.
        \label{eq:L-smoothness}
    \end{equation}
\end{assumption}
\begin{assumption}[Bounded variance and second moment] \label{assumption:bounded variance}
    For all $\mv \theta\in\mathbb{R}^{d\times 1}$, the stochastic gradient $\hat\nabla f_{k}(\mv \theta)$ is unbiased, and has bounded variance and second moment, i.e.,
    \begin{equation}
    \mathbb{E}\left[\left\|\hat\nabla f_{k}(\mv \theta)-\nabla f_{k}(\mv \theta)\right\|^2\right] \leq \sigma_l^2 \text{ and }
        \mathbb{E}\left[\left\|\hat{\nabla} f_{k}(\mv \theta)\right\|^2\right] \leq B^2_{k},
    \end{equation}
\end{assumption}
where the expectation $\mathbb{E}[\cdot]$ is taken over the choice of the mini-batch used to evaluate the stochastic gradient $\hat \nabla f_{k}(\mv \theta)$.
\begin{assumption}[Bounded Heterogeneity] \label{assumption:heterogeneity}
    For all $k \in [K]$ and $\mv \theta\in \mathbb{R}^{d\times 1}$, the gradients of the local empirical functions $f_{k}(\mv \theta)$ and of the global loss function $f(\mv \theta)=\frac{1}{K}\sum_{k=1}^{K}f_{k}(\mv \theta)$ satisfy the inequality
    \begin{equation}
        \left\|\nabla f_{k}(\mv \theta)-\nabla f(\mv \theta)\right\|^2 \leq \sigma_g^2.
    \end{equation}
\end{assumption}
These assumptions are standard and have been considered in prior art (see, e.g., \cite{basu19qsparse}). 
We also assume the following.
\begin{assumption}[i.i.d. channels] \label{assumption:iid channel}
    The channel coefficient $h_{k,j}^{(t)}$ are i.i.d. over the rounds $t=1,2,\ldots,T$ and symbols $j=1,2,\ldots,d$ for all devices $k=1,2,\ldots,K$. We denote as 
\begin{equation} \label{eq:lambda CCDF}
    \lambda_{k}\triangleq \text{Pr}(|h_{k,j}^{(t)}|^2> \epsilon_{k})
\end{equation}
the transmission probability for device $k$ when following the power control policy \eqref{eq:indicator variable}.
\end{assumption}





\subsection{Convergence Bound}
We study convergence in terms of the average gradient norm $\Myfrac{1}{T}\sum_{t=0}^{T-1} \mathbb{E}\| \nabla f(\boldsymbol\theta^{(t)}) \|^2$ as in, e.g., \cite{basu19qsparse}, and we prove the following theorem.
\begin{theorem} \label{theorem:convergence}
Under Assumptions 1-4, if the learning rate $\eta^{(t)}=\eta$ satisfies the inequality
\begin{equation}
    45\eta^3L^3Q^3+30\eta^2 Q^2 L^2+(\Myfrac{3}{2})\eta QL \le \Myfrac{1}{8},
    \label{eq:learning rate requirement}
\end{equation}
the expectation of the square gradient norm satisfies the inequality \eqref{eq:convergence bound} at the top of the next page, where the expectation $\mathbb{E}[\cdot]$ is over the small-scale channel fading coefficient $h_{k,j}^{(t)}$, the channel noise, and the stochastic gradients. In \eqref{eq:convergence bound}, we have defined the constants $B \triangleq \max_{k} B_{k}$, $L \triangleq \max_{k} L_{k}$, $C_{\lambda_{k}}=\Myfrac{4(1-\lambda_{k}^2)}{\lambda_{k}^2}$, and $\tilde{C}_{\lambda_k} = \Myfrac{4(1-\lambda_{k}^2)}{\lambda_{k}^2}+1$.

\begin{figure*}[t]
\begin{multline}
\frac{1}{T}\sum_{t=0}^{T-1} \mathbb{E}\| \nabla f(\boldsymbol\theta^{(t)}) \|^2 \le \underbrace{\frac{8}{\eta Q T}\left[f(\boldsymbol \theta^{(0)})-f^* \right]}_{\text{initilization error}}+ \underbrace{\frac{12}{K}\sum_{k=1}^{K}\eta^2 B^2 Q^2 L^2 C_{\lambda_{k}}}_{\text{contraction}}+\underbrace{\frac{8\eta L \sigma^2}{K^2}\max_{k \in [K]}\frac{\lambda_{k} B^2 Q \tilde{C}_{\lambda_k}}{P_{k} \kappa_{k} \epsilon_{k}}}_{\text{effective channel noise}} \\ 
+ \underbrace{(40\eta^2QL^2+60 \eta^3 Q^2L^3)(\sigma_l^2+6Q\sigma_g^2)+12\eta QL \sigma_l^2}_{\text{SGD \& data heterogeneity}}.
\label{eq:convergence bound}
\end{multline} 
\hrulefill
\end{figure*}
\end{theorem}
\begin{IEEEproof}
    {\color{black}A sketch of the proof can be found in Appendix \ref{subsec:proof of theorem}}. It relies on a perturbed iterate analysis that is widely used in the analysis for error-feedback-based SGD schemes (e.g., \cite{stich2018sparsified, basu19qsparse}).
\end{IEEEproof}
The bound in (\ref{eq:convergence bound}) takes into account the loss in learning performance in terms of convergence due to the randomness of the local SGD steps, of the deep fading channels, and of the channel noise.
By plugging in the learning rate $\eta=1/\sqrt{T}$, which satisfies condition \eqref{eq:learning rate requirement}, the bound \eqref{eq:convergence bound} indicates that AirFL-Mem converges to a stationary point at an average rate $\mathcal{O}(1/\sqrt{T})$. As a special case, by removing the effect of the fading channel, i.e., by setting $h_{k,j}^{(t)}=1$, and of the communication noise, i.e., by setting $\sigma^2 = 0$, the result in \eqref{eq:convergence bound} recovers the standard convergence rate $\mathcal{O}(1/\sqrt{T})$ of FedAvg \cite[Theorem 1]{yang2021achieving}.

Theorem \ref{theorem:convergence} can be used to bring insights into the role of the proposed long-term memory mechanism. 
To see this, consider AirFL with short-term memory, which uses the memory variable $\tilde{\mv{m}}_{k}^{(t)}=(\mv 1 - \mv q_{k}^{(t-1)}) \odot \mv \Delta_{k}^{(t-1)}$ to compensate for the distortion caused by truncation via \eqref{eq:transmitted signal of ECESA-DSGD} \cite{amiri2020federated}.  With this scheme, a counterpart of bound \eqref{eq:convergence bound} is derived in Appendix \ref{subsec:proof of short-term and without memory} (see \eqref{eq:bound of short-term}). With a learning rate $\eta = \Myfrac{1}{\sqrt{T}}$, this bound exhibits the same $\mathcal{O}(1/\sqrt{T})$ reduction of the average squared norm of the gradient, but it also demonstrates an error floor of order $\mathcal{O}((\Myfrac{B^2}{K})\sum_{k=1}^{K}(1-\lambda_k^2))$. The error floor is caused by deep fading, which causes the term $(1-\lambda_k^2)$, with \eqref{eq:lambda CCDF}, to be  larger than zero. In contrast, as mentioned, AirFL-Mem can converge to a stationary point with an arbitrarily small error. This shows that AirFL-Mem can successfully mitigate the impact of deep fading on the convergence of FL. 

\section{An Optimal Truncation-Threshold Design} \label{sec:optimization}
In this section, we leverage the bound in \eqref{eq:convergence bound} to introduce an  optimization strategy for the thresholds $\{\epsilon_k\}_{k=1}^{K}$ used in the power control policy \eqref{eq:indicator variable}. This optimization entails a  non-trivial trade-off, since increasing the threshold $\epsilon_k$ implies the transmission of fewer entries in the vector $\mv x_{k}^{(t)}$, while also increasing the power for each transmitted entry. Consequently, decreasing the threshold $\epsilon_k$ increases the probability of transmission, $\lambda_{k}$, while decreasing the power available for each transmitted entry.

Plugging $C_{\lambda_{k}}=\Myfrac{4(1-\lambda_{k}^2)}{\lambda_{k}^2}$ and $\tilde{C}_{\lambda_k} = \Myfrac{4(1-\lambda_{k}^2)}{\lambda_{k}^2}+1$ into \eqref{eq:convergence bound}, we formulate the problem of optimizing truncation-thresholds as the minimization of the bound \eqref{eq:convergence bound}, i.e.,
\begin{equation} \label{eq:original optimization problem}
    \begin{aligned}
        \mathrm{(P1)}: ~ &\mathop{\mathtt{Minimize}}_{\left\{\epsilon_{k}\right\}_{k=1,\ldots,K}} ~ \frac{1}{K}\sum_{k=1}^{K}48 \frac{1-\left(\lambda_{k}(\epsilon_k)\right)^2}{\left(\lambda_{k}(\epsilon_k)\right)^2}\eta^2 B^2 Q^2 L^2 \\
        & + \frac{8\eta L \sigma^2}{K^2}\max_{k \in [K]}\frac{\lambda_{k}(\epsilon_k) B^2 Q\left(\frac{4(1-\left(\lambda_{k}(\epsilon_k)\right)^2)}{\left(\lambda_{k}(\epsilon_k)\right)^2} + 1\right)}{P_{k} \kappa_{k} \epsilon_{k}} \\ 
    & \mathtt{Subject \ to} ~~~ \epsilon_{h}^{(i)}>0, \quad i=1,\ldots,K.
    \end{aligned} 
\end{equation}
Note that addressing problem \eqref{eq:original optimization problem} requires prior knowledge of the distribution of the channels.

We now study the practical case when the fading coefficients $h_{k,j}^{(t)}$, \(k\in[K]\), \(j\in[d]\), follow a circular symmetric complex Gaussian distribution with zero mean and unit variance, such that the transmission probability is $\lambda_{k} =\exp(-\epsilon_{k})$. With this choice, problem $\mathrm{(P1)}$ is reformulated as 
\begin{equation}
    \begin{aligned}
        & \mathrm{(P1^\prime)}:  ~\mathop{\mathtt{Minimize}}_{\left\{\lambda_{k}\right\}_{i=1,\ldots,K}}  ~ \frac{1}{K}\sum_{k=1}^{K}48 \frac{1-\lambda_{k}^2}{\lambda_{k}^2}\eta^2 B^2 Q^2 L^2 \\
        & \quad \quad \quad \quad  + \frac{8\eta L \sigma^2}{K^2}\max_{k \in [K]}\frac{\lambda_{k} B^2 Q\left(\frac{4(1-\lambda_{k}^2)}{\lambda_{k}^2} + 1\right)}{P_{k} \kappa_{k} \ln(1/\lambda_{k})}
     \\ 
    & \quad \quad \quad \mathtt{Subject \ to} ~~~ 0<\lambda_{k}<1, \quad i=1,\ldots,K.
    \end{aligned}
\end{equation}
\begin{proposition} \label{proposition:P2 convex}
    Problem $\mathrm{(P2)}$ is convex.
\end{proposition}
\begin{IEEEproof}
 See Appendix \ref{subsec:proof of P2 convex}.
\end{IEEEproof}
Since problem $\mathrm{(P1^\prime)}$ is convex, it can be optimally solved by interior point method \cite{boyd2004convex} via off-the-shelf software toolboxes such as CVX \cite{cvx}.

\section{Experiments}\label{sec:Experiments}

\begin{figure}[t] 
    \centering
    \hspace{-0.15in} \includegraphics[width=3.2in]{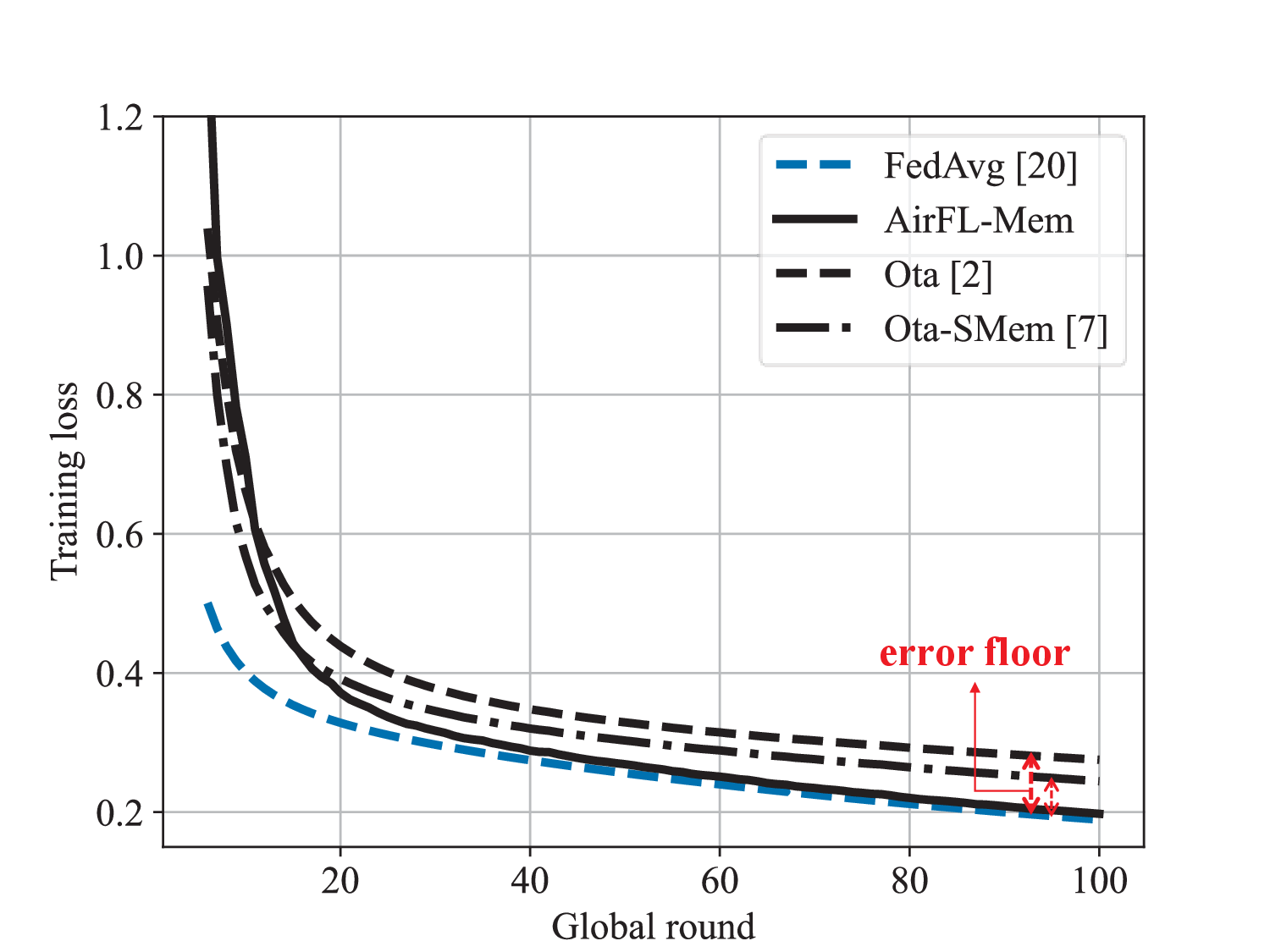}
    \includegraphics[width=3.15in]{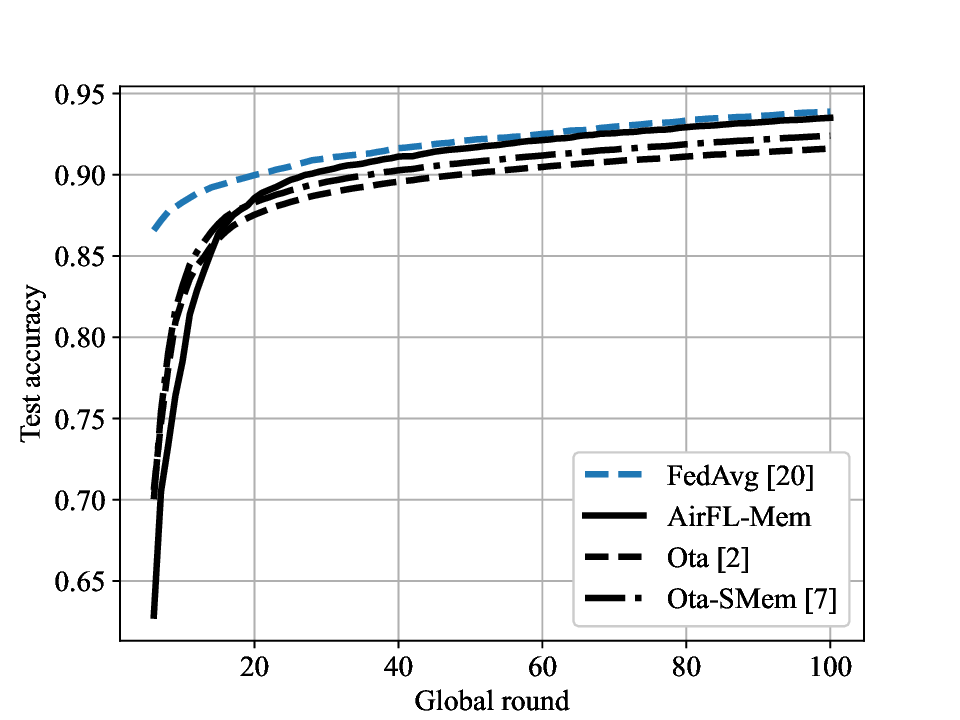}
    \vspace{-0.1in}
    \caption{Training loss and test accuracy versus the number of communication rounds, $T$.}
    \label{fig:loss_vs_T}
    \vspace{-0.1in}
\end{figure}


In this section, we evaluate the performance of AirFL-Mem in different setups with $K = 20$ devices, with the aim of comparing its performance with benchmarks, as well as of showing the effectiveness of the proposed  optimization of the truncation thresholds.

We consider the MNIST dataset learning task of classifying handwritten numbers, which is divided into $60,000$ training data samples and $10,000$ test data samples of $28\times28$ images. 
The samples are drawn randomly (without replacement) from the training set to form the local data set. All devices train a common DNN model that consists of one input layer with input shape $(28, 100)$, and one fully-connect layer of $100$ neurons, with ReLU activation function, and a softmax output layer, yielding a total number $d=79,510$  training parameters. SGD is adopted as an optimizer.

The simulation parameters are set as follows unless otherwise specified: the mini-batch size is $\vert\mathcal{D}_{k}^{(t)}\vert=64$; the number of local iterations $Q = 1$; the power constraint \(P_k = 2\times 10^{-6} \)\,W; the AWGN variance $\sigma^2=-83$\,dBm; the path loss $\kappa_{k}=\Myfrac{c^2}{\left(4\pi f_c r_{k}\right)^2}$, where $c$ is the speed of light, $f_c=2.4$\,GHz the central frequency, and $r_{k}$ the distance between device $k$ and edge server; the cell radius $100$\,m, in which the devices are uniformly located $r_{k}\sim \mathcal{U}(0,100)$ and all devices are involved in training; and Rayleigh fading channel is considered. To perform the truncation threshold optimization, we determine the parameters $B=0.1$ and $L=0.1$ via numerical grid search. 

We consider the following benchmarks: FedAvg with perfect communications \cite{mcmahan2017communication}; the truncated channel inversion-based vanilla AirFL scheme \cite{zhu2019broadband}, referred to as Ota; and AirFL with a short-term memory mechanism \cite{amiri2020federated}, which is referred to as Ota-SMem.
The thresholds of Ota and Ota-SMem benchmarks are set the same as AirFL-Mem after optimization.

We first compare the training loss and test accuracy of the AirFL-Mem and benchmarks versus the communication round $T$, in Fig. \ref{fig:loss_vs_T}. It is observed that the proposed AirFL-Mem approaches the ideal communication case, while Ota and Ota-SMem demonstrate significant error floors. 

\begin{figure}[t] 
    \centering
    \includegraphics[width=3.2in]{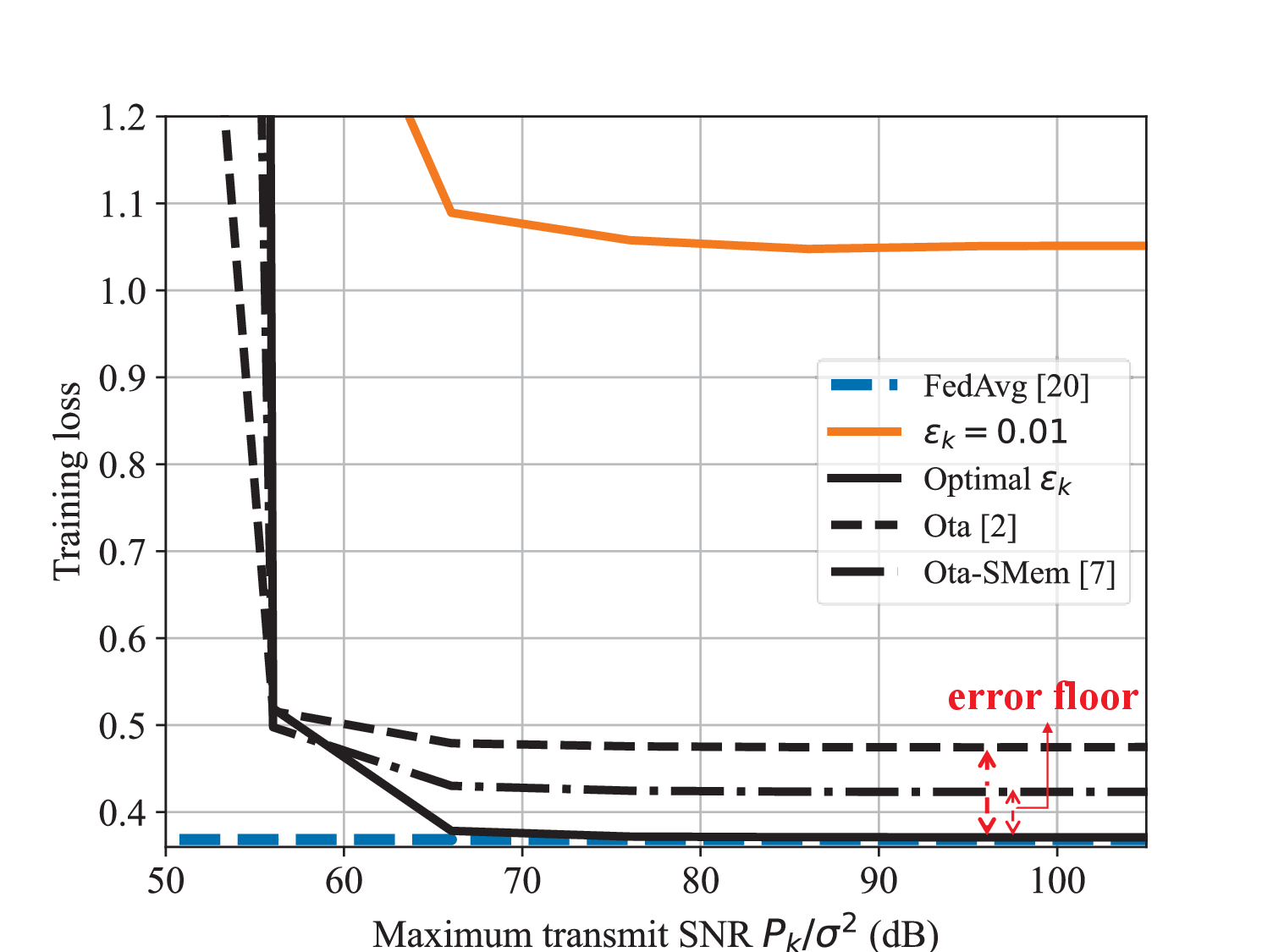}
    \vspace{-0.1in}
    \caption{Training loss  versus transmit SNR with $T=100$ iterations.}
    \label{fig:loss_vs_P}
    \vspace{-0.1in}
\end{figure}

Finally, we compare the training loss versus the maximum transmit signal-to-noise ratio (SNR) in Fig. \ref{fig:loss_vs_P}. For reference we also include the performance for a fixed threshold ($\epsilon=0.01$). 
The figure confirms that the proposed AirFL-Mem (with optimal thresholds) achieves performance close to FedAvg, as long as the SNR is large enough. In contrast,  Ota and Ota-SMem suffer from error floors even in the range of significantly high transmit SNR.

\section{Conclusions}
In this paper, we have proposed AirFL-Mem, an AirFL scheme that implements a  long-term memory mechanism to mitigate the impact of deep fading. For non-convex objectives, we have provided convergence bounds that suggest that AirFL-Mem enjoys the same convergence rate as FedAvg with ideal communications, while existing schemes exhibit error floors. The analysis was also leveraged to introduce a novel convex optimization scheme for the optimization of power control thresholds. Experimental results have demonstrated the effectiveness of the proposed approach.

\appendix
\subsection{Sketch of the proof of Theorem \ref{theorem:convergence}} \label{subsec:proof of theorem}
The proof of Theorem \ref{theorem:convergence} hinges on the fact that the channel-induced sparsification $\mv q^{(t)}_{k}\odot(\boldsymbol m^{(t)}_{k}+\boldsymbol \Delta^{(t)}_{k})$ satisfies the contraction property as \emph{rand-$k$} \cite{stich2018sparsified}, which is demonstrated by the following Lemma \ref{lemma:truncated rank-k contraction}. 

\begin{lemma}[Truncated channel inversion as rand-$k$ contraction] \label{lemma:truncated rank-k contraction}
    Under Assumption \ref{assumption:iid channel}, for all $\mv \theta \in \mathbb{R}^{d\times 1}$, we have, 
    \begin{equation}
        \mathbb{E}_{h}\left\| \mv \theta - \mv q_{k}^{(t)} \odot \mv \theta \right\|^2 = (1-\lambda_{k})\| \mv \theta \|^2,
    \end{equation}
    where $\mathbb{E}_{h}[\cdot]$ is over the small-scale channel fading $h_{k,j}^{(t)}$'s.
\end{lemma}
\begin{IEEEproof}
    See Appendix \ref{subsec:proof of contraction lemma}.
\end{IEEEproof}
With the aid of Lemma \ref{lemma:truncated rank-k contraction}, we  have the following memory bound \cite{basu19qsparse}.

\begin{lemma}[Bounded memory \cite{basu19qsparse}] \label{lemma:bounded memory}
    Under Assumption \ref{assumption:bounded variance} and \ref{assumption:iid channel}, for $\eta_t=\eta$, the expectation of the square norm of the long-term memory variable $\mv m_{k}^{(t)}$ for all $k\in[K]$ and $t\in [T]$, we have 
    \begin{equation}
        \mathbb{E}\| \mv m_{k}^{(t)} \|^2 \le \frac{4(1-\lambda_k^2)}{\lambda_k^2} \eta^2 B^2 Q^2.
    \end{equation}
\end{lemma}

Then we apply the perturbed iterate analysis as in \cite{stich2018sparsified} to provide the convergence bound of AirFL-Mem. Define the maintained virtual sequence $\{\tilde{\mv \theta}^{(t)}\}_{t=0,\ldots,T-1}$ as follows:
\begin{equation}
    \tilde{\mv \theta}^{(t+1)} = \tilde{\mv \theta}^{(t)}- \mv \Delta^{(t)} - \frac{\eta^{(t)}}{\sqrt{\rho^{(t)}}K}\mv n^{(t)},
\end{equation}
where $\mv \Delta^{(t)} = \frac{1}{K}\sum_{k=1}^{K}\mv \Delta_{k}^{(t)}$, and $\tilde{\mv \theta}^{(0)}=\mv \theta^{(0)}$. Then we have the following relations:
with $\mv{m}_{k}^{(t+1)}=\boldsymbol \Delta^{(t)}_{k}+\mv{m}_{k}^{(t)}-\boldsymbol q^{(t)}_{k}\odot (\boldsymbol \Delta^{(t)}_{k}+\boldsymbol m^{(t)}_{k})$ and $\mv{m}_{k}^{(0)}=\mv 0$, we have
\begin{equation}
    \mathbb{E}\left\| \mv{\theta}^{(t)} - \tilde{\mv \theta}^{(t)} \right\|^2 = \mathbb{E}\left\| \frac{1}{K}\sum_{k=1}^{K}\mv m_{k}^{(t)} \right\|^2 \le \frac{1}{K}\sum_{k=1}^{K}C_{\lambda_k} \eta^2 B^2 Q^2,
    \label{eq:virtual difference bound}
\end{equation}
where $C_{\lambda_k}=\Myfrac{4(1-\lambda_k^2)}{\lambda_k^2}$ and the inequality is followed by Jensen's inequality and Lemma \ref{lemma:bounded memory}.

The proof of convergence of AirFL-Mem begins with the $L$-smoothness of gradient (Assumption \ref{assumption: L-smoothness}). We first have
\begin{multline}
    f\left(\tilde{\mv \theta}^{(t+1)}\right) \le f\left(\tilde{\mv \theta}^{(t)}\right) - \left< \nabla f(\tilde{\mv \theta}^{(t)}), \mv \Delta^{(t)} + \frac{\eta^{(t)}}{\sqrt{\rho^{(t)}} K}\mv n^{(t)} \right> \\ 
    + \frac{L}{2}\left\| \mv \Delta^{(t)} + \frac{\eta^{(t)}}{\sqrt{\rho^{(t)}} K}\mv n^{(t)}\right\|^2.
\end{multline}
With some algebraic manipulations, for $\eta^{(t)}=\eta$, we arrive at \eqref{eq:per-round convergence} at the top of the next page.
\begin{figure*}[t]
    \begin{multline}
    \mathbb{E}\left[f\left( \tilde{\mv \theta}^{(t+1)} \right)\right] \le f\left(\tilde{\mv \theta}^{(t)}\right) - \frac{\eta Q}{4}\left\| \nabla f(\mv \theta^{(t)}) \right\|^2 + \frac{3\eta Q L^2}{2}\mathbb{E}\left\| \mv \theta^{(t)} - \tilde{\mv \theta}^{(t)} \right\|^2 + \frac{\eta L^2}{K}\sum_{k=1}^{K}\sum_{q=0}^{Q-1}\mathbb{E}\left\|\mv \theta_{k}^{(t,q)} - \mv \theta^{(t)} \right\|^2 + \frac{\eta^2 \sigma^2 Ld }{2K^2}\frac{1}{\rho^{(t)}} \\ 
    + \frac{3}{K}\sum_{k=1}^{K}\mathbb{E}\left\| \sum_{q=0}^{Q-1} \tilde{\nabla}f_{k}(\mv \theta_{k}^{(t,q)}) - \nabla f_{k}\left(\mv \theta_{k}^{(t,q)}\right) \right\|^2 + \frac{3}{K}\sum_{k=1}^{K}\mathbb{E}\left\| \nabla f_{k}\left(\mv \theta_{k}^{(t,q)}\right) - \nabla f_{k}(\mv \theta_{k}) \right\|^2 + 3Q^2\left\| \nabla f\left(\mv \theta^{(t)} \right) \right\|^2
    \label{eq:per-round convergence}
\end{multline}
\hrulefill
\end{figure*}
To bound $\Myfrac{1}{\rho^{(t)}}$, we introduce the following lemma. 
\begin{lemma}\label{lemma:bound of power scaling factor}
    Under Assumption \ref{assumption:bounded variance} and \ref{assumption:iid channel}, we have 
    \begin{equation}
        \frac{1}{\rho^{(t)}} \le \max_{k\in [K]} \left\{ \frac{2\lambda_k B^2 Q^2 (C_{\lambda_k}+1)}{d P_k \kappa_k \epsilon_k} \right\}.
        \label{eq:bound of power scaling factor}
    \end{equation}
\end{lemma}
\begin{IEEEproof}
    See Appendix \ref{subsec:proof of power scaling bound}.
\end{IEEEproof}

By applying the bound of $\mathbb{E}\| \mv{\theta}^{(t)} - \tilde{\mv \theta}^{(t)} \|^2$ in \eqref{eq:virtual difference bound}, Assumption \ref{assumption:bounded variance}-\ref{assumption:heterogeneity}, Lemma \ref{lemma:bounded memory}, Lemma 2 in \cite{yang2021achieving}, and Lemma \ref{lemma:bound of power scaling factor}, for $45\eta^3L^3Q^3+30\eta^2 Q^2 L^2+(\Myfrac{3}{2})\eta QL \le \Myfrac{1}{8}$, we arrive at the inequality \eqref{eq:simplified per-round convergence} at the top of the next page.
\begin{figure*}[t]
\begin{multline} \label{eq:simplified per-round convergence}
    \mathbb{E}\left[f\left( \tilde{\mv \theta}^{(t+1)} \right)\right] \le f\left(\tilde{\mv \theta}^{(t)}\right) - \frac{\eta Q}{8}\left\| \nabla f(\mv \theta^{(t)}) \right\|^2 + \frac{3\eta^3 Q^3 L^2 B^2}{2} \frac{1}{K}\sum_{k=1}^{K}C_{\lambda_k} + \left(5\eta^3 Q^2 L^2 + \frac{15}{2}\eta^4 Q^3L^3\right)(\sigma_l^2 + 6Q\sigma_g^2) \\
    + \frac{3\eta^2 L Q^2}{2}\sigma_l^2 + \frac{\eta^2 L}{K^2}\max_{k\in [K]} \left\{ \frac{\lambda_k B^2 Q^2 (C_{\lambda_k}+1)}{P_k \kappa_k \epsilon_k} \right\}
\end{multline}
\hrulefill
\end{figure*}
Performing a telescopic sum from $t =0$ to $T-1$, taking an average over all randomness, putting the $\mathbb{E}\| \nabla f(\mv \theta^{(t)})\|^2$ to the left, and rearranging at \eqref{eq:simplified per-round convergence}, we obtain the inequality \eqref{eq:convergence bound}, which completes the proof of Theorem \ref{theorem:convergence}.

\subsection{Convergence of AirFL with short-term memory and that without memory} \label{subsec:proof of short-term and without memory}
We first consider the convergence of AirFL with short-term memory. Note that the global update is given by \eqref{eq:airfl global update}. By the $L$-smoothness of gradients, we have
\begin{multline}
    f\left(\mv \theta^{(t+1)}\right) \le f\left(\mv \theta^{(t)}\right) \\ 
    - \left< \nabla f(\mv \theta^{(t)}), \frac{1}{K}\sum_{k=1}^{K}\mv g_{k}^{(t)} + \frac{\eta^{(t)}}{\sqrt{\rho^{(t)}} K}\mv n^{(t)} \right> \\ 
    + \frac{L}{2}\left\| \frac{1}{K}\sum_{k=1}^{K}\mv g_{k}^{(t)} + \frac{\eta^{(t)}}{\sqrt{\rho^{(t)}} K}\mv n^{(t)}\right\|^2.
\end{multline}
where $\mv g_{k}^{(t)} = \mv q_{k}^{(t)}\odot (\mv \Delta^{(t)}_{k}+\mv m^{\prime(t)}_{k})$. With some algebraic manipulations, we arrive at 
\begin{multline} \label{eq:per-round convergence short-term}
     \mathbb{E}\left[f\left( \mv \theta^{(t+1)} \right)\right] \le f\left(\mv \theta^{(t)}\right) - \frac{\eta Q}{2}\left\| \nabla f(\mv \theta^{(t)}) \right\|^2 \\ 
     + \frac{1}{\eta Q}\mathbb{E}\left\| \frac{1}{K}\sum_{k=1}^{K}\left(\mv g_{k}^{(t)} - \mv \Delta_{k}^{(t)} \right) \right\|^2   + \frac{L}{2K}\sum_{k=1}^{K}\mathbb{E}\| \mv g_{k}^{(t)} \|^2 \\ 
     + \frac{\eta L^2}{K}\sum_{k=1}^{K}\sum_{q=0}^{Q-1}\mathbb{E}\left\|\mv \theta_{k}^{(t,q)} - \mv \theta^{(t)} \right\|^2 + \frac{\eta^2 \sigma^2 Ld }{2K^2}\frac{1}{\rho^{(t)}},
\end{multline}
which has the similar structure as \eqref{eq:per-round convergence}. 
We bound the term $\mathbb{E}\left\| \mv g_{k}^{(t)} - \mv \Delta_{k}^{(t)} \right\|^2 \le 2(1-\lambda_k^2)\eta^2Q^2B^2$ using Assumption \ref{assumption:bounded variance} and \ref{assumption:iid channel}; bound the term $\mathbb{E}\|\mv \theta_{k}^{(t,q)} - \mv \theta^{(t)}\|^2$ using  Lemma 2 in \cite{yang2021achieving}; and bound the term $\mathbb{E}\| \mv g_{k}^{(t)} \|^2 \le 2 \lambda_k (2-\lambda_k)\eta^2 Q^2 B^2$ using Assumption \ref{assumption:bounded variance} and \ref{assumption:iid channel}. Moreover, the bound of $1/\rho^{(t)}$ can be easily obtained by the bound of $\mathbb{E}\| \mv g_{k}^{(t)} \|^2$ followed by the steps in Appendix \ref{subsec:proof of power scaling bound}.
Following the same strategy of Appendix \ref{subsec:proof of theorem}, for proper choice of $\eta$, we easily obtain the final convergence result as 
\begin{multline} \label{eq:bound of short-term}
    \frac{1}{T}\sum_{t=0}^{T-1} \mathbb{E}\| \nabla f(\boldsymbol\theta^{(t)}) \|^2 \le \underbrace{\frac{4}{\eta Q T}\left[f(\boldsymbol \theta^{(0)})-f^* \right]}_{\text{initilization error}} \\
    +\underbrace{\frac{8B^2}{K}\sum_{k=1}^{K}C_{\lambda_k}}_{\text{contraction}} +\underbrace{\frac{4\eta L \sigma^2}{K^2}\max_{k \in [K]}\frac{\lambda_k B^2 Q \tilde{C}_{\lambda_k}}{P_{k} \kappa_{k} \epsilon_{k}}}_{\text{effective channel noise}} \\ 
+ 20\eta^2L^2Q(\sigma_l^2+6Q\sigma_g^2)+\frac{4\eta Q B^2 L}{K}\sum_{k=1}^{K}\lambda_k\tilde{C}_{\lambda_k},
\end{multline}
where $C_{\lambda_k} = (1-\lambda_k^2)$ and $\tilde{C}_{\lambda_k} = (2-\lambda_k)$.
This completes the convergence proof of AirFL with short-term memory.

For the convergence of AirFL without memory, we only need to change $\mv g_{k}^{(t)}$ to $\mv g_{k}^{(t)} = \mv q_{k}^{(t)}\odot \mv m^{\prime(t)}_{k}$ and follow the same steps, and then we can easily obtain the convergence result of AirFL without memory. Finally, we will obtain the same form as \eqref{eq:bound of short-term} with $C_{\lambda_k} = (1-\lambda_k)$ and $\tilde{C}_{\lambda_k} = 1/2$.

\subsection{Proof of Lemma \ref{lemma:truncated rank-k contraction}} \label{subsec:proof of contraction lemma}
For all $\mv \theta \in \mathbb{R}^{d \times 1}$, we have
\begin{equation}
    \begin{aligned}
        & \mathbb{E}_{h}\left\| \mv \theta - \mv q_{k}^{(t)} \odot \mv \theta \right\|^2 \\ 
        & = \mathbb{E}_{h}\left\| \mv \theta  \right\|^2 - 2\mathbb{E}_{h}\left< \mv \theta, \mv q_{k}^{(t)} \odot \mv \theta  \right> + \mathbb{E}_{h}\left\| \mv q_{k}^{(t)} \odot \mv \theta \right\|^2 \\ 
        & \overset{(a)}{=} \mathbb{E}_{h}\left\| \mv \theta  \right\|^2 - \mathbb{E}_{h}\left\| \mv q_{k}^{(t)} \odot \mv \theta \right\|^2,
        \label{eq:proof of contraction 1}
    \end{aligned}
\end{equation}
where (a) is due to $\left< \mv \theta, \mv q_{k}^{(t)} \odot \mv \theta  \right> = \| \mv q_{k}^{(t)} \odot \mv \theta \|^2$. Then we calculate the term $\mathbb{E}_{h}\| \mv q_{k}^{(t)} \odot \mv \theta \|^2$ as follows.
\begin{equation}
    \begin{aligned}
        \mathbb{E}_{h}\left\| \mv q_{k}^{(t)} \odot \mv \theta \right\|^2
        & = \mathbb{E}_{h}\left[ \sum_{m=1}^{d}|\theta_m|^2 q_{k,j}^{(t)} \right] \\ 
        & = \sum_{m=1}^{d}\left( |\theta_m|^2\mathbb{E}_{h}\left[q_{k,j}^{(t)}\right]\right) = \lambda_{k} \|\mv \theta \|^2,
        \label{eq:proof of contraction 2}
    \end{aligned}
\end{equation}
which yields the desired result.

\subsection{Proof of Lemma \ref{lemma:bound of power scaling factor}} \label{subsec:proof of power scaling bound}
Record that the power constraint for all $k\in [K]$ is given by 
\begin{equation}
    \frac{1}{d}\mathbb{E} \left\| \mv p_{k}^{(t)} \odot \mv q_{k}^{(t)} \odot \boldsymbol x_{k}^{(t)} \right\|^2 \le P_{k},
\end{equation}
where $\mv x_{k}^{(t)} = (\boldsymbol \Delta^{(t)}_{k} + \boldsymbol m^{(t)}_{k})/\eta^{(t)}$ for AirFL-Mem. 
By \eqref{eq:indicator variable}, we can obtain 
\begin{equation}
\begin{aligned}
    & \mathbb{E}\left\| \mv p_{k}^{(t)} \odot \mv q_{k}^{(t)} \odot \boldsymbol x_{k}^{(t)} \right\|^2 \\
    & = \frac{\rho^{(t)}}{\kappa_k \eta^2} \mathbb{E}\left[ \sum_{j=1}^{d}\left( \frac{q_{k,j}^{(t)}}{|h_{k,j}^{(t)}|^2} \left(  \Delta^{(t)}_{k,j} +  m^{(t)}_{k,j} \right)^2 \right) \right] \\ 
    & \le \frac{\rho^{(t)}}{\kappa_k \epsilon_k \eta^2} \mathbb{E}\left[  \sum_{j=1}^{d}\left( q_{k,j}^{(t)} \left(  \Delta^{(t)}_{k,j} +  m^{(t)}_{k,j} \right)^2 \right) \right] \\ 
    & = \frac{\rho^{(t)}}{\kappa_k \epsilon_k \eta^2} \sum_{j=1}^{d}\left( \mathbb{E}[q_{k,j}^{(t)}] \mathbb{E}\left[ \left(  \Delta^{(t)}_{k,j} +  m^{(t)}_{k,j} \right)^2 \right] \right) \\ 
    & = \frac{\rho^{(t)} \lambda_k }{\kappa_k \epsilon_k \eta^2} \mathbb{E}\| \mv \Delta^{(t)}_{k} + \mv m^{(t)}_{k} \|^2 \\ 
    & \le \frac{\rho^{(t)} \lambda_k }{\kappa_k \epsilon_k \eta^2} (2\mathbb{E}\| \mv \Delta^{(t)}_{k} \|^2 +  2\mathbb{E}\|\mv m^{(t)}_{k} \|^2) \\ 
    & \le \underbrace{\frac{\rho^{(t)} \lambda_k }{\kappa_k \epsilon_k}2 B^2Q^2(C_{\lambda_k}+1)}_{A(\rho^{(t)})}.
\end{aligned}
\end{equation}
where the last inequality is followed by Assumption \ref{assumption:bounded variance} and Lemma \ref{lemma:bounded memory}.
We choose $\tilde{\rho}$ such that $A(\tilde{\rho}) \le dP_k$ for all $k\in [K]$ is satisfied, i.e., 
\begin{equation}
    \tilde{\rho} \triangleq \min_{k\in [K]} \frac{d P_k \kappa_k \epsilon_k }{2\lambda_k B^2Q^2(C_{\lambda_k}+1)}.
\end{equation}
By the long-term power constraint and the definition of $\tilde{\rho}$, we have $\rho^{(t)} \ge \tilde{\rho}$, i.e., 
\begin{equation}
    \frac{1}{\rho^{(t)}} \le \frac{1}{\tilde{\rho}} = \max_{k\in [K]} \frac{2\lambda_k B^2Q^2(C_{\lambda_k}+1)}{d P_k \kappa_k \epsilon_k},
\end{equation}
which yields the desired result.

\subsection{Proof of Proposition \ref{proposition:P2 convex}} \label{subsec:proof of P2 convex}
Define functions $g(x)=\Myfrac{a\left(1-x^2\right)}{x^2}$ and $f_{k}(x)=\Myfrac{b_{k}\left(\Myfrac{4\left(1-x^2\right)}{x}+x\right)}{\log \left(\frac{1}{x}\right)}$, where $a$ and $b_{k}$ are some constants, and $\text{dom}f=\text{dom}g=(0,1)$. The function $f(x)$ is convex since
\begin{multline}
    f_{k}^{\prime \prime}(x)
    =\frac{b_{k}\left(\frac{8\left(\log ^2\left(\frac{1}{x}\right)+1\right)}{x^3}+\left(-\frac{12}{x^3}-\frac{3}{x}\right) \log \left(\frac{1}{x}\right)-\frac{6}{x}\right)}{\log ^3\left(\frac{1}{x}\right)}
\end{multline}
is positive in $0<x<1$.
And $g(x)$ is convex since $g^{\prime\prime}(x) = \Myfrac{6a}{x^4} > 0$ for all $0<x<1$.
Then we use the fact that the finite sum of convex functions is convex and $\max_{k} f_{k}$ is convex when $f_{k}$ is convex, which yields the objective in Problem $\mathrm{(P2)}$ is convex in $(0,1)$.

\addtolength{\topmargin}{-0.01in}
\bibliographystyle{IEEEtran}
\bibliography{DL_ref}

\end{document}